\documentclass[11pt]{article}
\usepackage{amsmath,amssymb}

\usepackage{float}
\usepackage{graphicx}
\usepackage{wrapfig}
\usepackage{slashed}
\usepackage{hyperref}
\usepackage{placeins}
\usepackage{cite}
\usepackage{longtable}
\usepackage{xcolor}
\usepackage{geometry}
\usepackage{subcaption}
\def\beq{\begin{eqnarray}}
\def\eeq{\end{eqnarray}}

\def\la{\langle }
\def\ra{\rangle }

\newcommand{\be}{\begin{equation}}
\newcommand{\ee}{\end{equation}}
\newcommand{\bea}{\begin{eqnarray}}
\newcommand{\eea}{\end{eqnarray}}
\newcommand{\bg}{\begin{gather}}

\newcommand{\bseq}{\begin{subequations}}
\newcommand{\eseq}{\end{subequations}}

\newcommand{\bra}[1]{\langle #1 |}
\newcommand{\ket}[1]{| #1 \rangle}

\def\be{\begin{eqnarray}}
\def\ee{\end{eqnarray}}

\newcommand{\CK}{{\mathcal{K}}}
\newcommand{\CH}{{\mathcal{H}}}

\usepackage{booktabs}
\begin{document}

\title{\textbf{Average Spread Complexity and \\ the Higher-Order Level Spacing}}
\vspace{2cm}
\date{}
\author{ \textbf{  Amin Faraji Astaneh$^{a,b}$ and  Niloofar Vardian$^b$ }} 
\maketitle
\begin{center}
\hspace{-0mm}
\emph{$^a$ 
Department of Physics, Sharif University of Technology,\\
P.O.Box 11155-9161, Tehran, Iran}
 \end{center}
\begin{center}
  \hspace{-0mm}
   \emph{ $^b$  Research Center for High Energy Physics,\\
Department of Physics, Sharif University of Technology,\\
P.O.Box 11155-9161, Tehran, Iran} \\
\end{center}
%%%%%%%%%%%%%%%%%%%%%%%%%%%%%
\begin{abstract}
\noindent {
We investigate the spread complexity of a generic two-level subsystem of a larger system to analyze the influence of energy level statistics, comparing chaotic and integrable systems. Initially focusing on the nearest-neighbor level spacing, we observe the characteristic slope-dip-ramp-plateau structure. Further investigation reveals that certain matrix models exhibit additional iterative peaks, motivating us to generalize the known spacing distributions to higher-order level spacings. While this structure persists in chaotic systems, we find that integrable systems can also display similar features, highlighting limitations in using complexity as a universal diagnostic of quantum chaos.
}
\end{abstract}

\vskip 8 cm
\noindent
\rule{7.7 cm}{.5 pt}\\
\noindent 
\noindent
\noindent ~~~ {\footnotesize e-mails:\ faraji@sharif.ir ,  niloofar.vardian72@sharif.edu}

\newpage
\section{Introduction}
The concept of complexity may seem intuitive and straightforward at first glance, but accurately defining it within a quantum system is quite challenging. This difficulty has led to a unique development in high-energy physics. Typically, a concept is first understood in quantum mechanics, then explored in the continuum limit of quantum field theory, and potentially examined later through gauge/gravity duality in holographic gravitational duals. In this context, specific holographic proposals for complexity were initially presented in AdS/CFT, drawing from black hole physics, even though the foundational aspects of these concepts in quantum mechanics were not yet fully established.
Two of the most prominent conjectures in holographic complexity are the $\text{Complexity}=\text{Volume}$ (CV) and $\text{Complexity}=\text{Action}$ (CA) conjectures. The CV conjecture proposes that the complexity of a quantum state in the boundary theory is dual to the volume of a minimal spacelike hypersurface anchored to the boundary (or equivalently, in the black hole context, the volume of the Einstein-Rosen bridge) \cite{susskind2016computational,stanford2014complexity}. The CA conjecture, on the other hand, establishes a duality between complexity and the gravitational action evaluated on the Wheeler-de Witt (WdW) patch \cite{brown2015complexity,brown2016complexity,lehner2016gravitational}.

As stated before, while these holographic conjectures have provided valuable insights, it is crucial to establish a precise and microscopically rigorous definition of complexity within quantum field theory.

However, recent developments have introduced novel bottom-up approaches to complexity, which are now under active investigation.
A recently proposed and rigorously defined notion of complexity is Krylov complexity, first introduced in \cite{parker2019universal}.

The central idea in this context is to map the evolution of a Heisenberg operator onto a one-dimensional chain, representing the minimal orthonormal basis that captures the operator’s dynamics. As time progresses, the operator’s growth is reflected in its movement along this chain, with the average position defining the Krylov complexity, which can also be interpreted as the operator's size. 

This framework has recently been extended to describe quantum states undergoing unitary evolution and spreading throughout Hilbert space. In this context, the spread complexity, which is analogous to the Krylov complexity for quantum states, quantifies the average position of the state’s spread along a one-dimensional chain.

One of the most significant aspects of investigating Krylov complexity, as highlighted in recent studies, lies in its potential to yield deeper insights and more precise classifications of quantum chaos.

In this context, Krylov complexity has been extensively studied in a wide range of quantum chaotic systems, including random matrix theory (RMT) \cite{balasubramanian2022quantum, balasubramanian2025quantum, tang2023operator, caputa2024krylov, bhattacharjee2025krylov}, quantum billiards \cite{hashimoto2023krylov, camargo2024spectral, balasubramanian2024chaos}, quantum spin chains \cite{rabinovici2022krylov, scialchi2024integrability, gill2024complexity, bhattacharya2024krylov, camargo2024spread, scialchi2025exploring}, various versions of the Sachdev-Ye-Kitaev (SYK) model \cite{rabinovici2021operator, bhattacharjee2023krylov, hornedal2022ultimate, erdmenger2023universal, chapman2024krylov, baggioli2024krylov} and some aspects of many body quantum systems and conformal field theories \cite{Nandy:2023brt, Banerjee:2022ime}. For a review that compiles many of these studies, see also \cite{Baiguera:2025dkc} and \cite{Nandy:2024htc}.

Its applications span diverse areas, such as topological and quantum phase transitions \cite{caputa2022quantum, afrasiar2023time, caputa2023spread, pal2023time, Chakrabarti:2025hsb}, quantum batteries \cite{kim2022operator}, high-energy quantum field theory \cite{caputa2024krylov,Alishahiha:2022anw, Vardian:2024fsp, Iizuka:2023pov, Iizuka:2023fba, Malvimat:2024vhr, Vasli:2023syq, Camargo:2022rnt, Kundu:2023hbk, Adhikari:2022whf, Avdoshkin:2022xuw, Dymarsky:2021bjq}, thermalization \cite{Alishahiha:2024rwm}, bosonic systems modeling ultra-cold atoms \cite{bhattacharyya2023operator}, saddle-dominated scrambling \cite{bhattacharjee2022krylov, huh2024spread}, and open quantum systems \cite{bhattacharya2022operator, bhattacharjee2023operator, mohan2023krylov, bhattacharya2023krylov, bhattacharjee2024operator, carolan2024operator}.

In particular, in \cite{balasubramanian2022quantum}, spread complexity is examined in chaotic systems, including the SYK model and various random matrix models. It has been found that the spread complexity for thermofield double (TFD) states exhibits four distinct dynamical regimes: a linearly increasing ramp that reaches a peak, followed by a decline to a plateau. Based on these findings, it has been suggested that the behavior of spread complexity resembles the pattern of the spectral form factor (SFF) in chaotic systems, which also follows a slope-dip-ramp-plateau structure \cite{guhr1998random, cotler2017black}.

Building on this, the authors of \cite{erdmenger2023universal} demonstrated that a peak before saturation in the spread complexity of the TFD states is a universal characteristic of chaotic systems, see also\cite{Bhattacharyya:2023grv}. They argued that this late-time peak in spread complexity is directly related to the ramp phase of the SFF. Since the ramp in the SFF arises from spectral rigidity and level spacing statistics, this suggests that the peak prior to saturation in spread complexity is a distinctive feature of quantum chaos. This observation has been supported by numerical studies across various models \cite{huh2024spread,bhattacharjee2025krylov, balasubramanian2024chaos}. For example, in studies such as \cite{huh2024krylov, baggioli2024krylov, camargo2024spread}, the authors investigated the late-time saturation of spread complexity during phase transitions from integrable to maximally chaotic regimes. Their findings consistently revealed that the peak before saturation becomes sharper as the system approaches maximal chaos, while it is entirely absent in integrable systems.

These results highlight the role of spread complexity as an effective diagnostic tool for identifying quantum chaos. It was further demonstrated in \cite{alishahiha2024krylov} that, based on level spacing statistics, integrable systems exhibit complexity that approaches saturation from below. In contrast, chaotic systems, particularly Gibbs states and TFD states, reach saturation only after exceeding a distinct peak.

It is important to note that, unlike classical chaos, quantum chaos does not have a universally accepted definition. Nevertheless, the most widely recognized criterion for quantum chaos is based on the statistical properties of energy level spacings. Therefore, it is reasonable to think that the distribution of level spacings plays a crucial role in the emergence of a peak in spread complexity for chaotic systems. 
Chaotic systems are often modeled using RMT, where the level spacings typically follow the Wigner-Dyson distribution. In the context of the TFD state, all energy levels contribute to the spread complexity. However, for certain initial states, only a subset of energy levels may have a significant impact on the spread complexity. Therefore, the higher-order level spacing may play a significant role.

To be more concrete, let us consider a generic initial state in a $d$-level quantum system, expressed in the energy eigenbasis as
\begin{equation}\label{inist}
    \ket{\psi} = \sum_{i=1}^d c_i \ket{E_i}\, ,
\end{equation}
where $\{\ket{E_i}\}$ are the Hamiltonian's eigenstates with energies $\{E_i\}$. The time evolution of $\ket{\psi}$ is confined to the subspace spanned by eigenstates with nonvanishing amplitudes $c_i \neq 0$. Explicitly, defining
\begin{equation}
    \mathcal{H}_\psi = \text{span}\{ \ket{E_\alpha} \mid c_\alpha \neq 0 \}\, ,
\end{equation}
the time-evolved state $\ket{\psi(t)} = e^{-iHt}\ket{\psi}$ always satisfies $\ket{\psi(t)} \in \mathcal{H}_\psi$.  

This dynamics can be equivalently described by the projected Hamiltonian  
\begin{equation}
    H_\psi = P_\psi H P_\psi\, ,
\end{equation}
where $P_\psi$ projects onto $\mathcal{H}_\psi$. The spectrum of $H_\psi$ consists of the original energies $\{E_i\}$ for which $\langle E_i | \psi \rangle \neq 0$, and thus it depends on the initial state's support.  

The spread complexity of $\ket{\psi(t)}$ is intimately tied to the level statistics of $H_\psi$. To see this, let us sort the original Hamiltonian's spectrum as $E_1 < \dots < E_d$. The projected spectrum $\{\widetilde{E}_\alpha\}$ of $H_\psi$ corresponds to a subset $\{E_{f(\alpha)}\}$ labeled by an increasing function $f(\alpha)$, where $\alpha = 1, \dots, d_\psi = \dim(\mathcal{H}_\psi)$.  

The $n$th order level spacing, $s^{(n)} \equiv E_{i+n} - E_i$ allows us to express the spacings of $H_\psi$ as  
\begin{equation}
    \widetilde{s}_\alpha \equiv \widetilde{E}_{\alpha+1} - \widetilde{E}_\alpha = s^{(n)}_\alpha, \quad n = f(\alpha+1) - f(\alpha)\, .
\end{equation}
This connection motivates studying how spread complexity reflects the underlying level statistics; Poissonian for integrable systems versus Wigner-Dyson for chaotic systems.  

As a case study, we consider a superposition state ($d_\psi = 2$)\footnote{We use this terminology throughout the text: by superposition state, we mean to select two kets from a $d$-level system and create a superposition. We also refer to this as a two-level subsystem.}. We observe that increasing the energy gap $\widetilde{s} = \widetilde{E}_2 - \widetilde{E}_1$ (i.e., probing higher-order spacings in the full spectrum) sharpens the peak in spread complexity, with distinct signatures for integrable and chaotic spectra. If we continue increasing this difference, we notice that new dips and peaks emerge before reaching saturation. This behavior is also observed in integrable systems. Therefore, subsectors of an integrable system can exhibit chaotic characteristics, meaning this feature alone cannot serve as a reliable indicator of quantum chaos. See \cite{Jeong:2024oao} for other discussions on behaviors earlier suggested for Krylov complexity in chaotic systems, which is also seen in some integrable toy models.

The paper is organized as follows: In Section 2, we review the concepts of spread complexity and the Lanczos algorithm. In Section 3, we calculate the spread complexity for a superposition state. Section 4 presents the results for the average spread complexity of a system with a generic Wigner distribution. In Section 5, we investigate the average spread complexity of a subsystem in relation to the higher level spacing of the original system, considering both chaotic and integrable systems. Finally, we conclude our work in the concluding section.

To facilitate clarity and quick reference, we list the following abbreviations used in this manuscript:

\vspace{.5 cm}
\begin{tabular}{@{}ll@{}}
\textbf{RMT} & Random Matrix Theory\\
\textbf{TFD} & Thermofield Double (state) \\
\textbf{SYK} & Sachdev-Ye-Kitaev (model) \\
\textbf{GOE} & Gaussian Orthogonal Ensemble \\
\textbf{GUE} & Gaussian Unitary Ensemble \\
\textbf{GSE} & Gaussian Symplectic Ensemble \\
\textbf{SFF} & Spectral Form Factor \\
\end{tabular}
\section{A Conscious Review on the Lanczos Algorithm and the Spread Complexity}

Consider the time evolution of a generic quantum state under the successive application of the Hamiltonian $H$. The state at time $t$ can be expressed as
\be
\vert \psi_t\rangle = \sum_{n=0}^\infty \frac{(-it)^n}{n!} \vert \psi_n\rangle \, , \quad \text{where} \quad \vert \psi_n\rangle = H^n \vert \psi_0\rangle \, .
\ee
Here, $\vert \psi_0\rangle$ is the initial state, and $\vert \psi_n\rangle$ represents the state after $n$ applications of the Hamiltonian.

The Lanczos algorithm is essentially a Gram-Schmidt orthogonalization procedure that constructs an orthonormal basis from the sequence of states $\{H^n \vert \psi_0 \rangle\}$. This orthonormal basis is known as the Krylov basis, denoted by $\{\vert K_n\rangle\}$.

To construct the Krylov basis, the algorithm proceeds as follows
\begin{itemize}
    \item[i)] Start with the initial state $\vert K_0\rangle = \vert \psi_0\rangle$.
    \item[ii)] Generate the next state $\vert k_1\rangle$ by applying the Hamiltonian and subtracting the projection onto the previous state
    \be
    \vert k_1\rangle = H\vert K_0\rangle - \langle K_0\vert H \vert K_0\rangle \vert K_0\rangle \, .
    \ee
    Normalize $\vert k_1\rangle$ to obtain $\vert K_1\rangle$
    \be
    \vert K_1\rangle = \frac{1}{\| k_1\|} \vert k_1\rangle \, .
    \ee
    \item[iii)] Repeat this process iteratively to generate the entire Krylov basis. The recursive relation for generating the basis is given by
    \be
    \vert k_{n+1}\rangle = (H - a_n)\vert K_n\rangle - b_n \vert K_{n-1}\rangle \, ,
    \ee
    where $\vert K_n\rangle = b_n^{-1} \vert k_n\rangle$. The coefficients $a_n$ and $b_n$ are defined as
    \be
    a_n \equiv \langle K_n \vert H \vert K_n \rangle \, , \quad b_n \equiv \| k_n \| \, ,
    \ee
    with the initial condition $b_0 = 0$. These coefficients are known as the \emph{Lanczos coefficients}.
\end{itemize}

This iterative procedure terminates after a finite number of steps, resulting in a Krylov basis that spans a $d_\CK$-dimensional subspace $\CK$ of the $d_\CH$-dimensional Hilbert space $\CH$. The state vector at any time can be expanded in this basis as
\be
\vert \psi_t\rangle = \sum_{n=0}^{d_\CK-1} \phi_n(t) \vert K_n\rangle \, .
\ee
The Schrödinger equation then yields
\be\label{schrodinger}
i \partial_t \phi_m(t) = a_m \phi_m(t) + b_{m+1} \phi_{m+1}(t) + b_m \phi_{m-1}(t) \, .
\ee

Intuitively, a more complex state resulting from time evolution requires a greater number of Krylov blocks to accurately describe and construct it. This principle forms the basis for defining complexity. This idea leads us to the following definition of complexity, known as Krylov complexity or, when referring to states, spread complexity.
\be
C_\psi=\sum n |\phi_n(t)|^2=\sum n |\la\psi(t)|K_n\ra |^2\, .
\ee
The choice of the expansion coefficient is based on the idea that the cost function, represented by the spread complexity, can be interpreted as the average position of a particle in the Krylov quantum chain as time evolves.

To enhance computational efficiency, it is advantageous to introduce the return amplitude\footnote{Also referred to as the survival amplitude, the Loschmidt amplitude, and the auto-correlation function.} alongside the return probability as follows
\begin{equation}
    R(t)= \langle \psi_t \ket{\psi_0}=\phi_0^*\ , \ \mathcal{P}(t) = |R(t)|^2\, .
\end{equation}

If we take the initial state to be a thermofield double state
\begin{equation}
    \ket{\psi_0} = \frac{1}{\sqrt{Z(\beta)}} \sum_n e^{- \beta E_n /2} \ket{n} \otimes \ket{n}\, ,
\end{equation}
where $ Z(\beta) = \sum_n e^{- \beta E_n}$ is the partition function, the return probability reads
\begin{equation}
    \mathcal{P}(t) =  \frac{1}{Z(\beta)^2} \sum_{m,n} e^{-\beta (E_m+E_n)} e^{i(E_m-E_n)t}\, ,
\end{equation}
which is nothing but the spectral form factor (SFF) defined as \cite{guhr1998random, brezin1997spectral}
\begin{equation}
    \text{SFF}(t) = \frac{|Z(\beta + it )|^2}{|Z(\beta )|^2}\, . 
\end{equation}
The SFF is a powerful tool for probing the dynamics of quantum chaos, which is closely tied to level statistics. In chaotic systems, such as those described by RMT, the SFF exhibits a characteristic slope-dip-ramp-plateau behavior.
Moreover, in the case of $\beta =0$ where we have the maximally mixed state, the late time behavior of Krylov complexity is related to the infinite time average of the SFF \cite{del2017scrambling, baggioli2024krylov}
\begin{equation}
    \lim _{T \rightarrow \infty} \frac{1}{T} \int _0 ^ T \text{SFF} (t) = \frac{1}{ 1+ 2 C_\psi(t\rightarrow\infty)}
\end{equation}
while $ C_\psi(t\rightarrow\infty) = \frac{d-1}{2}$ where $d$ is the size of the system.

Now, we address the question of how to algorithmically calculate the Lanczos coefficients and, consequently, the complexity, given the return amplitude. The starting point involves calculating the following moments
\begin{equation}
M_n = i^n \left. \frac{d^n R(t)}{dt^n} \right|_{t=0} = (-1)^n \langle K_0 | H^n | K_0 \rangle , .
\end{equation}
The first few terms are given by:
\begin{equation}
M_1 = -\la K_0|H|K_0\ra=-a_0\  , \ M_2=\la K_0|H^2|K_0\ra = a_0^2 + b_1^2\ , \ M_3=-a_0^3-(2a_0+a_1)b_1^2\, ,
\end{equation}
and this procedure can be continued to obtain all the coefficients up to the desired order. In most practical situation, the moments $M_n$ are efficiently calculable and the task is to compute the Lanczos coefficients out of them.
\section{Spread Complexity of a Superposition State}
Let us start with a generic $d$-level quantum system with Hamiltonian 
\begin{equation}
 H = \sum _i E_i \ket{E_i} \bra{E_i}\, ,   
\end{equation}
where $\{E_i\}$ and $ \{\ket{E_i}\}$ are the set of eigenvalues and eigenfunctions, respectively. We assume that
\begin{equation}\label{sort}
    E_1 < E_2 < \dots < E_d\, .
\end{equation}
In a simple setup, we pick two eigenstates and construct a pure superposition state
\begin{equation}\label{state}
    \ket{\psi_0} = c_i \ket{E_i} + c_j \ket{E_j} = \cos \theta \ket{E_i} + \sin \theta e^{i \gamma} \ket{E_j}\, ,
\end{equation}
with $ 0 < \theta < \pi / 2 $ and $ 0 \leq \gamma \leq \pi$. 

The return amplitude for the state is 
\begin{equation}\label{SA}
    R(t) = \langle \psi_t | \psi_0 \rangle = e^{i E_i t} (\cos ^2 \theta + \sin ^2 \theta e^{-i s_{ij} t})\, ,
\end{equation}
where $s_{ij} \equiv E_i - E_j$.

Utilizing the algorithm mentioned in the previous section and by calculating the set of moments, one can determine the two sets of Lanczos coefficients. The moments simply read
\begin{equation}
M_n = (-1)^n \left( E_i^n \cos^2 \theta + E_j^n \sin^2 \theta \right)\, .
\end{equation}
Then the non-zero Lanczos coefficients will be determined as 
\begin{equation}
    \begin{split}
        & a_0 = \frac{1}{2}(E_i + E_j) + \frac{1}{2} s_{ij} \cos (2 \theta)\, ,\\
        & a_1 = \frac{1}{2}(E_i + E_j) - \frac{1}{2} s_{ij} \cos (2 \theta)\, ,
    \end{split}
\end{equation}
and
\begin{equation}
    \begin{split}
        & b_0 = 1\, ,\\
        & b_1 = \frac{1}{2} s_{ij} \sin (2 \theta)\, .
    \end{split}
\end{equation}
Now utilizing \eqref{schrodinger}, one obtains 
\begin{equation}\label{phi}
    \begin{split}
        & \phi_0 = R^*(t) = e^{-i E_i t} (\cos ^2 \theta + \sin ^2 \theta e^{i s_{ij} t})\, ,\\
        & \phi_1 = \frac{1}{2} e^{-i E_i t} (1 - e^{i s_{ij} t}) \sin(2 \theta)\, ,\\
        & \phi_{n > 1} = 0\, .
    \end{split}
\end{equation}
Clearly, $ |\phi_n|^2$ are functions of $s_{ij}$, and thus the spread complexity depends only on the difference of the energies. Finally, the spread complexity of the state is determined as
\begin{equation}\label{spread2}
    C_\psi(t, \theta) = \sin ^2 \left( \frac{s_{ij} t}{2} \right) \sin^2 (2 \theta)\, .
\end{equation}
We observe that the pattern of spread complexity as a function of time remains the same for different values of $\theta$. Furthermore, 
\begin{equation}
    C_\psi(\theta) = C_\psi(\pi/2 - \theta)\, .
\end{equation}
Thus, for $\frac{|c_i|}{|c_j|} = r$ or $\frac{1}{r}$, we obtain the same result, and the spread complexity depends only on $|c_i c_j|$. The maximum value is achieved when $|c_i| = |c_j| = \frac{1}{\sqrt{2}}$ so the complexity reaches its maximum at $\theta = \pi / 4$, which corresponds to the maximally mixed state. 

Additionally, it is important to note that for $\pi/4 \leq \theta < \pi/2$, the state admits an effective TFD description, where
\begin{equation}
    \beta = \frac{2}{s_{ij}} \log \tan \theta\, ,
\end{equation}
for the two-level Hamiltonian $H_{ij} = P_{ij} H P_{ij}$ 
with $P_{ij}$ being the projection operator onto the subspace $V_{ij} = \text{span} \{ \ket{E_i}, \ket{E_j}\}$.
\section{Average Spread Complexity for a Two-Level Subsystem with Wigner-Dyson Distribution}
In the previous section, we selected two energy levels to create a pure superposition state. However, we did not specify which energy levels were chosen from the spectrum. The concept of level spacing has been introduced to address this important question.

We firstly define the nearest neighbor level spacing as 
\begin{equation}
    s_i = E_{i+1} -E_i\, ,
\end{equation}
where $E_i$ represent the eigenvalues of the Hamiltonian of system sorted as in \eqref{sort}. The values $s_i$ could span a wide range; they might be very small, very large, or comparable to the mean level spacing $\langle s \rangle$. It is crucial to investigate the distribution of the nearest neighbor level spacing, $P(s)$. This distribution reveals where the energy eigenstates in a system's spectrum are more densely concentrated and where they are more sparsely distributed. Such insights are essential for determining whether a system exhibits integrability or chaos, which are fundamental to characterizing its dynamics.

For a very important case, the RMT, the level spacing is given by Wigner surmise. Although the Wigner surmise originates from the study of $2 \times 2$ matrices, it works good enough for matrices of higher dimensions. The Wigner surmise is given by
\begin{equation}
    P_{\alpha}(s) = C(\alpha) s^\alpha \exp( - A(\alpha) s^2)\, ,
\end{equation}
where $\alpha$ will be explained shortly. The constants $C(\alpha)$ and $A(\alpha)$ are determined by imposing the normalization conditions
\begin{equation}\label{normal}
     \int_0^\infty P(s)~ ds =\int_0^\infty s P(s)~ ds = 1 \, .
\end{equation}
The first condition ensures that $P(s)$ is a valid probability distribution, stating that the total probability of all possible outcomes must sum to one. The second condition ensures that the mean level spacing $\la s \ra$
is equal to one. In the context of RMT, it is indeed conventional to rescale the energy levels such that the average spacing between adjacent levels is unity. This rescaling simplifies the analysis and allows for universal comparisons between different systems.
Solving these conditions yields
\begin{equation}\label{AC}
    A(\alpha) = \left[ \frac{\Gamma\left(\frac{\alpha+2}{2}\right)}{\Gamma\left(\frac{\alpha+1}{2}\right)} \right]^2, \quad 
    C(\alpha) = \frac{2 \, A(\alpha)^{\frac{\alpha+1}{2}}}{\Gamma\left(\frac{\alpha+1}{2}\right)} \,,
\end{equation}
where $\Gamma(z) = \int_0^\infty t^{z-1} e^{-t} dt $ is the Gamma function.

Now, let us discuss what the parameter $\alpha$ tracks. It is well-known that random matrices depend primarily on the symmetry class of the matrices rather than their microscopic details. The parameter $\alpha$ labels these symmetry classes. Three well-known and particularly important classes are as follows

\begin{itemize}
    \item Systems with time-reversal symmetry are described by matrices belonging to the \emph{Gaussian Orthogonal Ensemble} (GOE).
    \item Systems with spin-rotational symmetry that break time-reversal symmetry are described by matrices belonging to the \emph{Gaussian Unitary Ensemble} (GUE).
    \item Systems with time-reversal symmetry but broken spin-rotational symmetry are described by matrices belonging to the \emph{Gaussian Symplectic Ensemble} (GSE).
\end{itemize}

\begin{figure}[h!]
\centering
        \includegraphics[width=.6\linewidth]{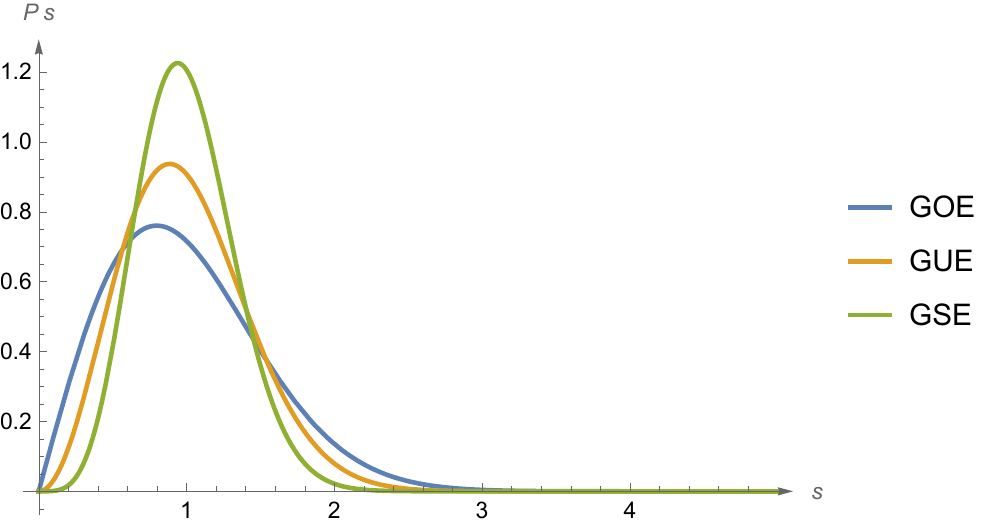}
    \caption{ Distribution of nearest neighbor level spacings $ P(E_{i+1} - E_i)$ for RMT.}
    \label{fig1}
\end{figure}

In the Wigner surmise, the parameter $\alpha$ takes the values $\alpha = 1$, $\alpha = 2$, and $\alpha = 4$ for GOE, GUE, and GSE, respectively. Correspondingly, the probability distributions for these three well-known classes are given by
\be
P_\alpha(s) = 
\begin{cases} 
    \dfrac{\pi}{2} \, s \, \exp\left(-\dfrac{\pi}{4} s^2\right), & \alpha = 1, \quad \text{GOE} 
    \\[1.5em]
    \dfrac{32}{\pi^2} \, s^2 \, \exp\left(-\dfrac{4}{\pi} s^2\right), & \alpha = 2, \quad \text{GUE} 
    \\[1.5em]
    \dfrac{2^{18}}{3^6 \pi^3} \, s^4 \, \exp\left(-\dfrac{64}{9\pi} s^2\right), & \alpha = 4, \quad \text{GSE} 
\end{cases}
\ee
and plot \eqref{fig1} is to schematically visualize their behavior

Although the number of these classes with universal characteristics is limited, as we will explain further, the other values of alpha, which eventually span a continuum spectrum, would also be meaningful. This motivates us to introduce the \emph{average spread complexity}\footnote{A distinct approach is implemented in \cite{Craps:2024suj}, where the system is initialized via a multi-seed configuration. This constitutes a fundamentally different physical framework from the one we examine.}. For a two dimensional RMT with a generic Wigner surmise distribution we have 
\begin{equation}\label{Ckbar}
        \bar{C}_{\psi}(t, \theta) = \int ds~ P_\alpha (s) C_\psi(t,\theta)=  \frac{1}{2} \left[ 1 - {}_1F_1\left(\frac{\alpha + 1}{2}; \frac{1}{2}; -\frac{t^2}{4A(\alpha)}\right) \right]\sin ^2 (2 \theta)\, ,
\end{equation}
where ${}_1F_1(a;b;z)$ is the confluent hypergeometric function.

\begin{figure}[h!]
        \centering
        \includegraphics[width=0.6\linewidth]{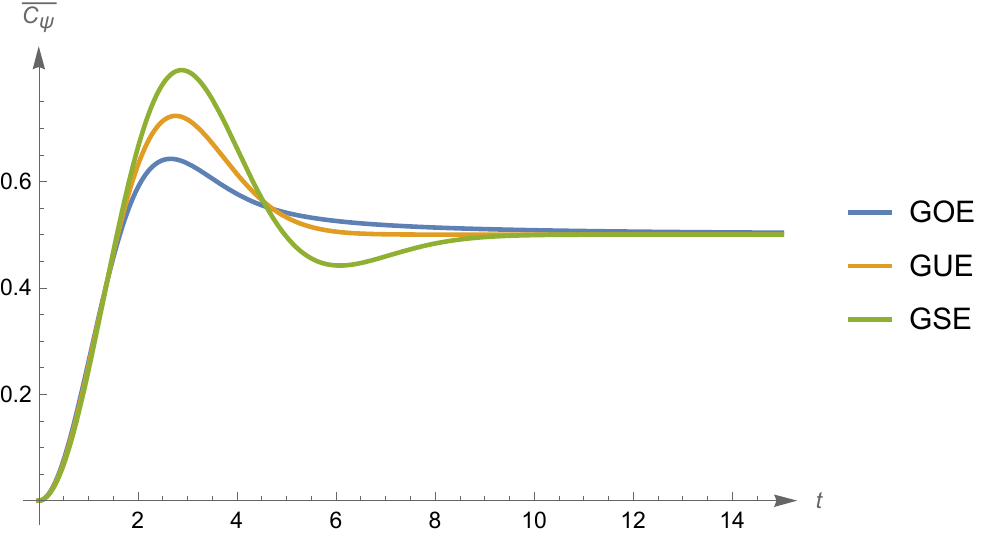}
    \caption{ Averaged spread complexity of the two dimensional random matrix theory for the maximally mixed states.}
    \label{fig2}
\end{figure}

\begin{figure}[h!]
    \centering
    % First subfigure
    \begin{subfigure}[b]{0.48\textwidth}
        \centering
        \includegraphics[width=\linewidth]{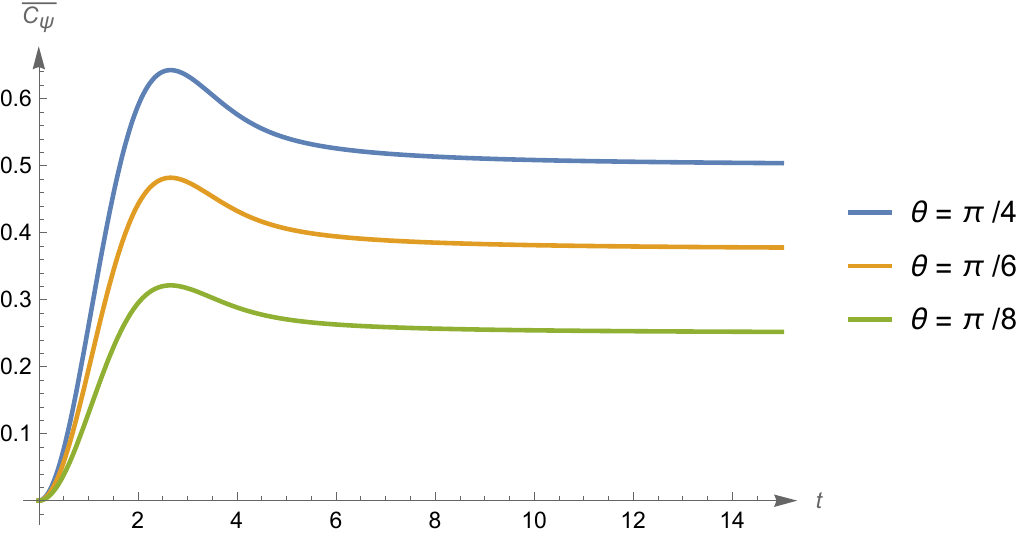}
        \caption{GOE}
        \label{fig:subfig1}
    \end{subfigure}
    \hfill
    % Second subfigure
    \begin{subfigure}[b]{0.48\textwidth}
        \centering
        \includegraphics[width=\linewidth]{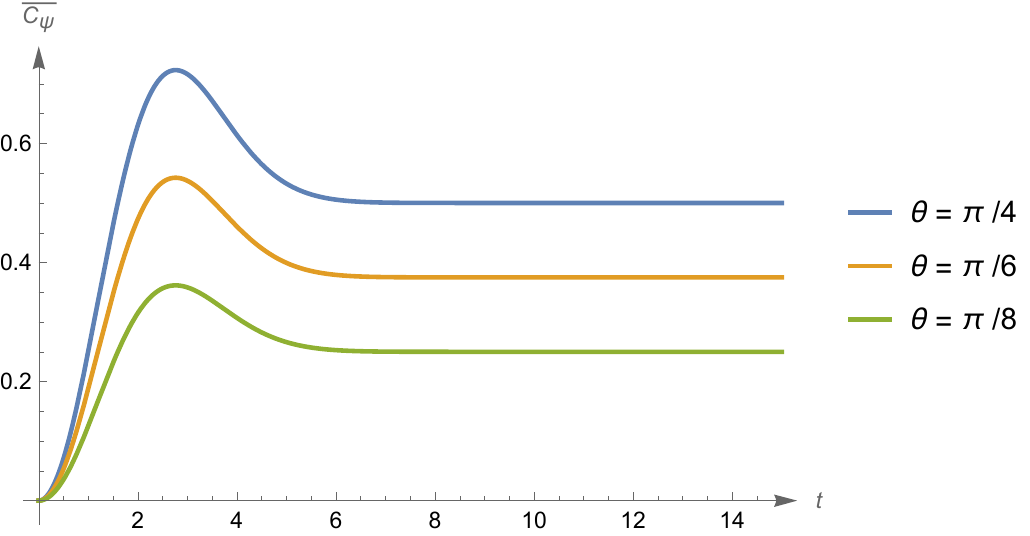}
        \caption{GUE}
        \label{fig:subfig2}
    \end{subfigure}
    
    % Third subfigure
    \begin{subfigure}[b]{0.48\textwidth}
        \centering
        \includegraphics[width=\linewidth]{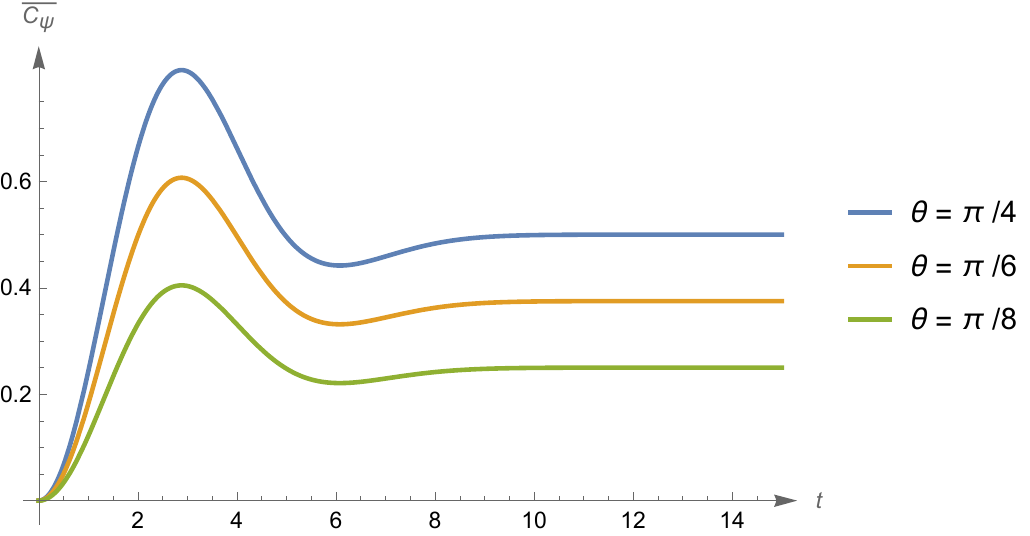}
        \caption{GSE}
        \label{fig:subfig3}
    \end{subfigure}
    
    \caption{Average spread complexity of 2-dimensional RMT for different initial states}
    \label{fig3}
\end{figure}

In the case of Wigner-Dyson distribution one obtains
\begin{equation}\label{averageCk}
  \bar{C}_{\psi}(t, \theta) = 
  \begin{cases}
    \dfrac{t}{\sqrt{\pi}} \, D\left(\dfrac{t}{\sqrt{\pi}}\right) \sin^2(2\theta), & \alpha = 1, \quad \text{GOE}
    \\[1.5em]
    \dfrac{1}{2} \left[1 +\frac{1}{8}\exp\left(-\dfrac{\pi t^2}{16}\right) \left(\pi t^2-8\right)\right] \sin^2(2\theta), & \alpha = 2, \quad \text{GUE}
    \\[1.5em]
    \dfrac{1}{2} \left[1 - \exp\left(-\dfrac{9\pi t^2}{256}\right) \left(1 + \dfrac{9\pi t^2}{16384} \left(3\pi t^2-256\right)\right)\right] \sin^2(2\theta), & \alpha = 4, \quad \text{GSE}
  \end{cases}
\end{equation}
where $F(z)$ is Dawson function defined as 
\begin{equation}
    D(z) = e^{-z^2} \int_0^z e^{u^2} ~du = \frac{\sqrt{\pi}}{2} e^{-z^2} \text{erfi(z)},
\end{equation}
while $\text{erfi(z)}$
is the imaginary error function. In the following figures, we present the average spread complexity for various two-dimensional random matrix theories. In \eqref{fig2}, we have plotted the average complexity for all three distributions for a maximally mixed state, allowing for a direct comparison in a single figure. In \eqref{fig3}, we separately examine the behavior for different values of $\theta$ for each of the three distributions individually.
\\
It is worth mentioning, as indicated in equation \eqref{averageCk}, that the saturation value for all three distributions (and, in general, for all Wigner distributions) is
\begin{equation}
    \bar{C}_\psi(t\rightarrow \infty,\theta) = \frac{1}{2} \sin ^2(2 \theta).
\end{equation}
With this understanding in mind, we will now connect these findings by providing a detailed explanation of the average complexity behavior based on the analysis of the presented plots.

In all three random matrix theories, a distinct peak can be observed before reaching saturation. As the parameter $\alpha$ increases, the first peak becomes sharper when transitioning from the GOE to the GSE. Additionally, in the GSE, a small dip appears prior to achieving saturation. This behavior has also been observed in TFD states within higher-dimensional random matrix theory. \cite{balasubramanian2022quantum}. 

\begin{figure}[h!]
        \centering
        \includegraphics[width=.6\linewidth]{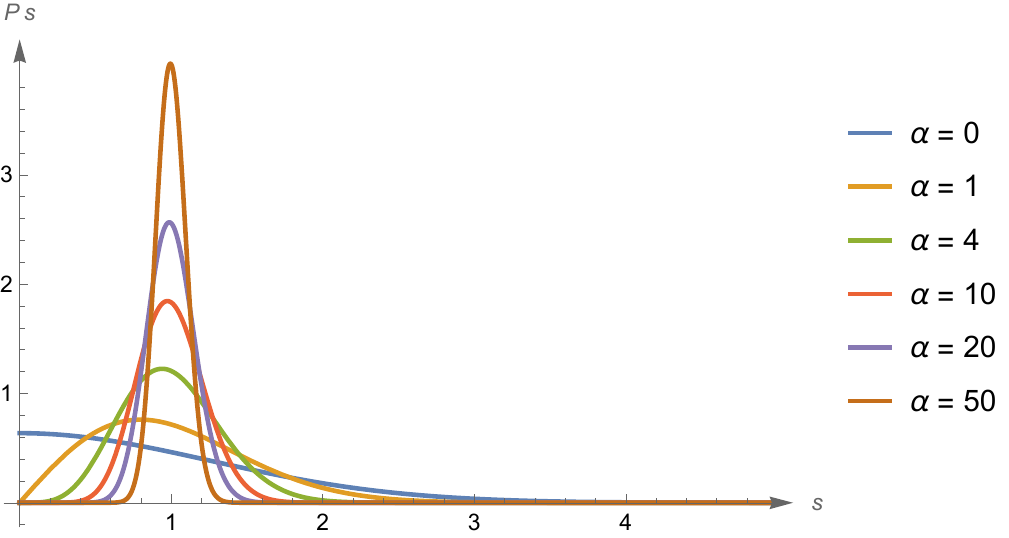}
    \caption{Wigner surmise distribution for different $\alpha$.}
    \label{fig4}
\end{figure}

\begin{figure}[h!]
        \centering
        \includegraphics[width=.6\linewidth]{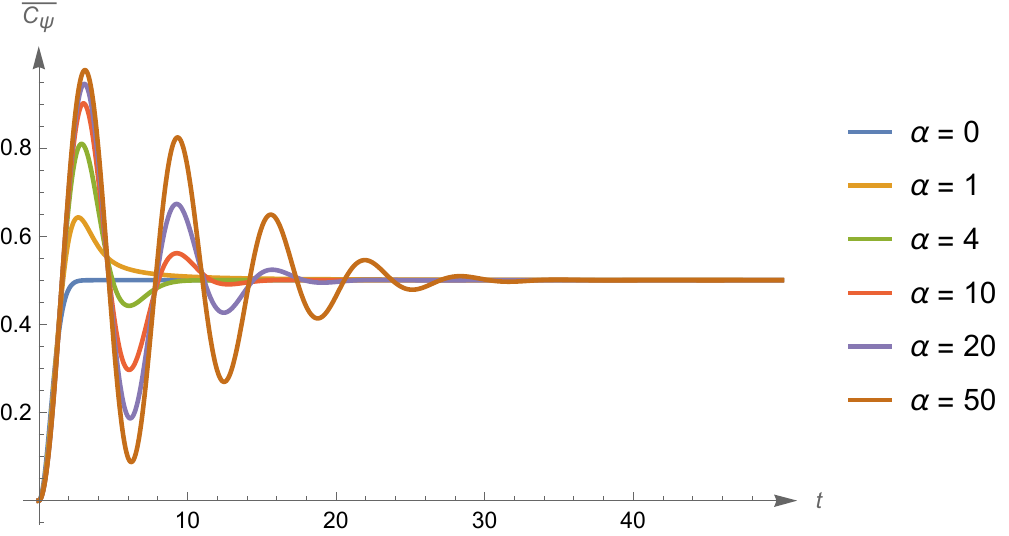}
    \caption{Average spread complexity of a superposition state for generic Wigner distribution. }
    \label{fig5}
\end{figure}

\begin{figure}[h!]
        \centering
        \includegraphics[width=.6\linewidth]{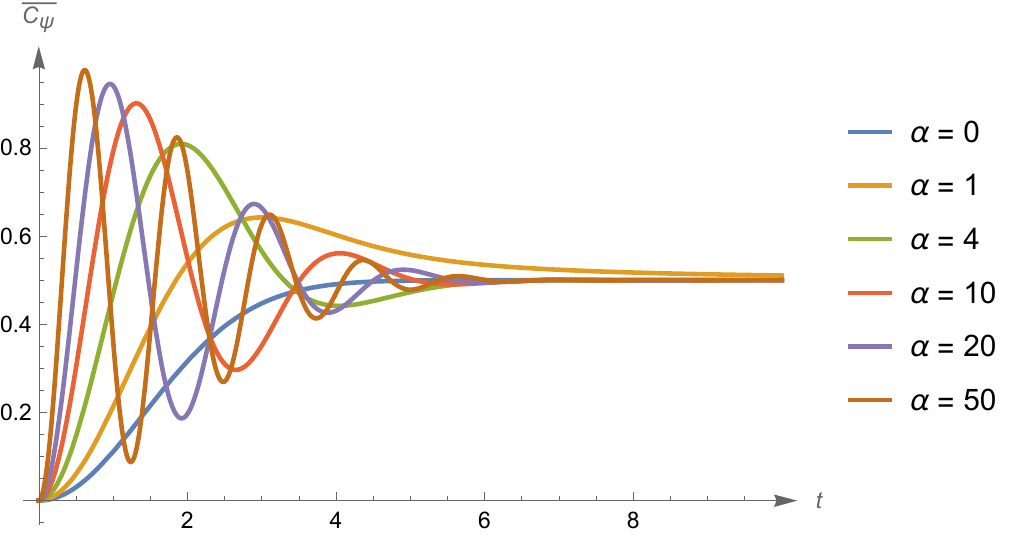}
    \caption{Average spread complexity of a superposition state for generic non- normalized Wigner-Dyson distribution. }
    \label{fig6}
\end{figure}

\begin{figure}[h!]
    \centering
    % First subfigure
    \begin{subfigure}[b]{0.6\textwidth}
        \centering
        \includegraphics[width=\linewidth]{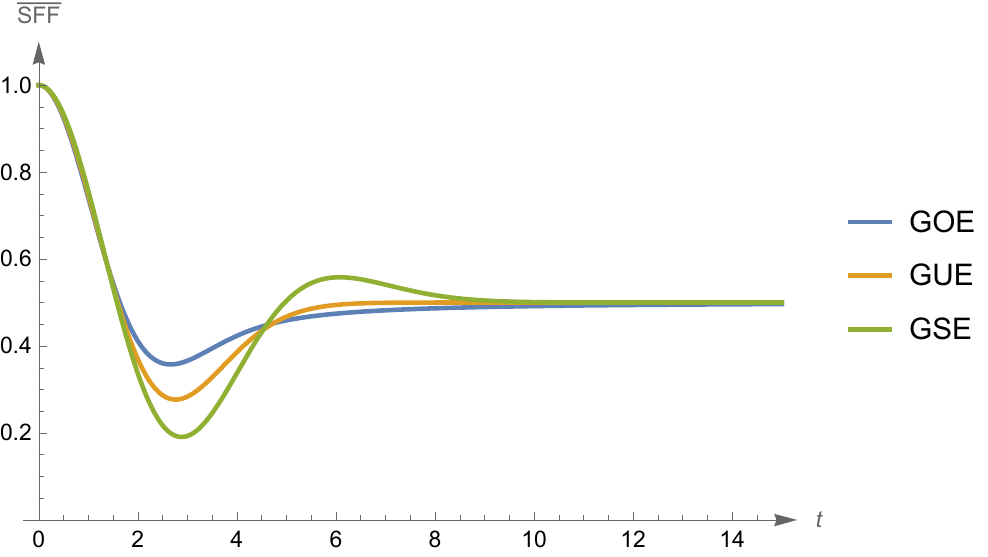}
        \caption{Wigner-Dyson distribution }
        \label{fig:subfig1}
    \end{subfigure}
    \hfill
    % Second subfigure
    \begin{subfigure}[b]{0.6\textwidth}
        \centering
        \includegraphics[width=\linewidth]{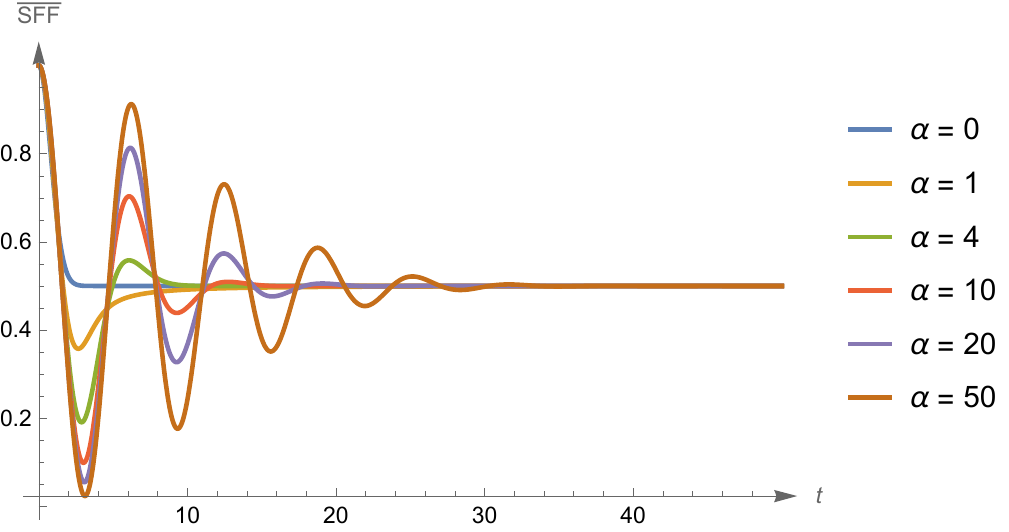}
        \caption{Wigner distribution for different $\alpha$.}
        \label{fig:subfig2}
    \end{subfigure}
    \caption{Average SFF of two dimensional subsystem for $\beta =0$ and different Wigner distribution}
    \label{fig7}
\end{figure}

\begin{figure}[h!]
    \centering
    % First subfigure
    \begin{subfigure}[b]{0.48\textwidth}
        \centering
        \includegraphics[width=\linewidth]{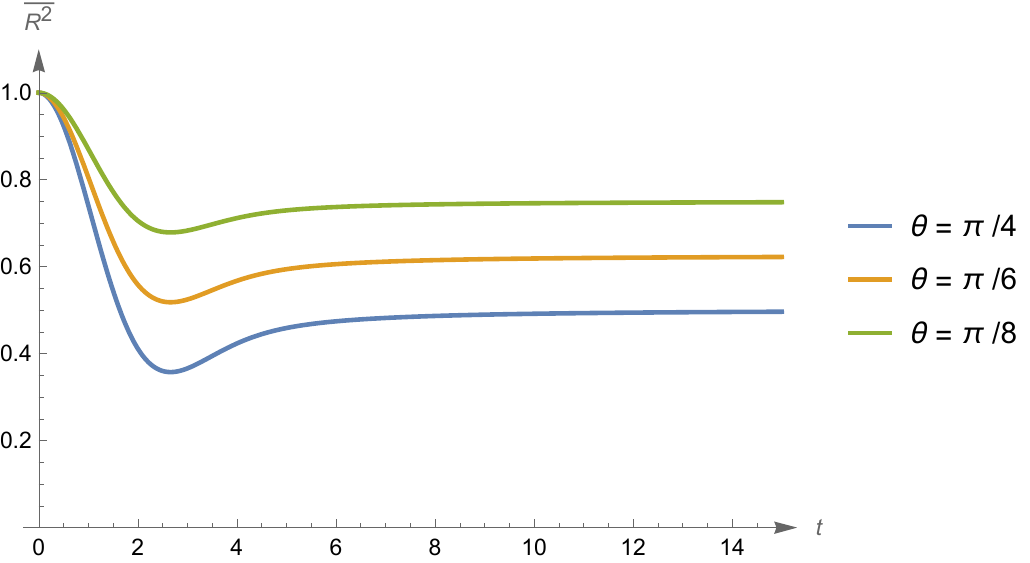}
        \caption{GOE}
        \label{fig:subfig1}
    \end{subfigure}
    \hfill
    % Second subfigure
    \begin{subfigure}[b]{0.48\textwidth}
        \centering
        \includegraphics[width=\linewidth]{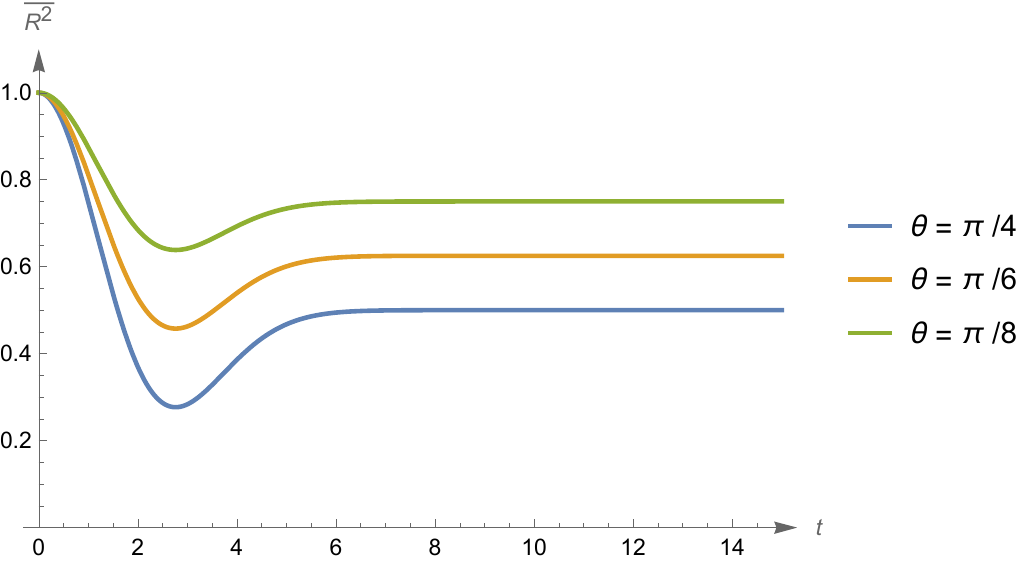}
        \caption{GUE}
        \label{fig:subfig2}
    \end{subfigure}
     % third subfigure
    \begin{subfigure}[b]{0.48\textwidth}
        \centering
        \includegraphics[width=\linewidth]{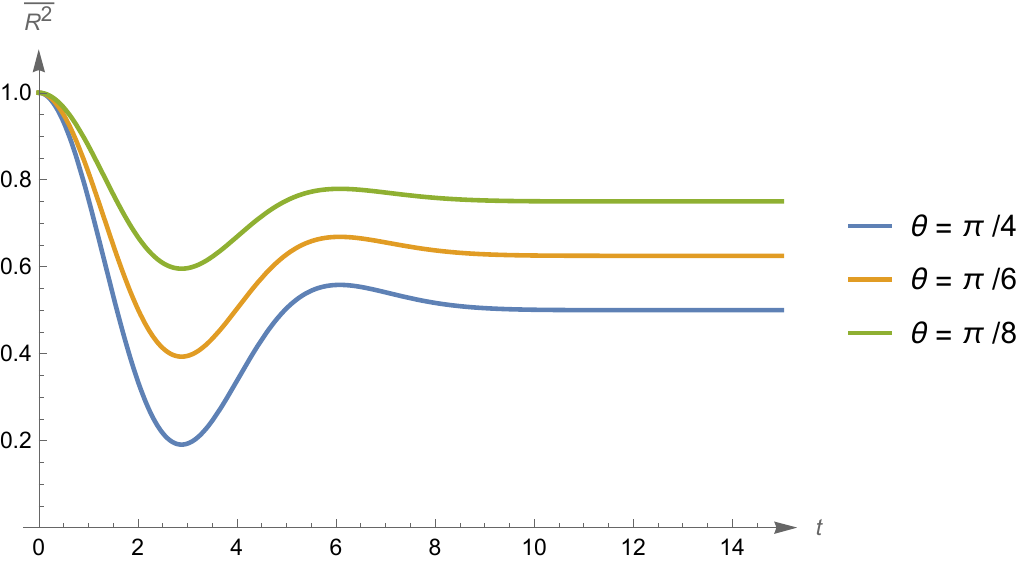}
        \caption{GSE}
        \label{fig:subfig2}
    \end{subfigure}
    \caption{Average square of survival amplitude of two dimensional RMT for different initial states}
    \label{fig8}
\end{figure}

Furthermore, it is widely discussed in the literature \cite{balasubramanian2022quantum, erdmenger2023universal, camargo2024spread, alishahiha2024krylov, baggioli2024krylov, camargo2024higher, huh2024krylov} that this behavior is a common feature of chaotic systems. Specifically, for TFD states in chaotic systems, the spread complexity is expected to reach saturation after a peak, following a period of early growth.
Let us fix $\theta$ and examine what happens as we increase the value of $\alpha$. In plots \eqref{fig4} and \ref{fig5}, we can observe the Wigner distribution and the average spread complexity for the maximally mixed state at different values of $\alpha$, respectively. 
As $\alpha$ increases, the first peak becomes sharper. Additionally, the number of peaks and dips before the average spread complexity reaches saturation increases. 
It is evident that as $\alpha$ grows, it takes longer for the system to reach saturation. However, if we consider the non-normalized distribution given by the equation $P_\alpha(s) \propto s^\alpha e^{-s^2}$ we find that the number of peaks and dips remains the same, but the time to saturation is consistent across different values of $\alpha$, as shown in plot \eqref{fig6}.
We will soon provide a compelling physical interpretation for this trend toward higher values of $\alpha$.
Before that, let us take a closer look at the behavior of the plots we have depicted.
As argued in \cite{balasubramanian2022quantum, erdmenger2023universal}, the presence of a peak before saturation is linked to the ramp in the SFF or the return probability of the TFD state. It is thus natural to study the average return probability of an initial state, first for a two-dimensional RMT model, and more generally, for systems with a generic Wigner eigenvalue distribution.
From \eqref{SA}, we simply obtain
\begin{equation}
    \mathcal{P}(t, \theta) = |R(t, \theta)|^2 
    = \sin^4 \theta + \cos^4 \theta + 2 \sin^2 \theta \cos^2 \theta \cos(s t) 
    = \text{SFF}\left(t, \beta = \frac{2}{s} \log \tan \theta\right)\,.
\end{equation}
The average $\mathcal{P}(t, \theta)$ for a generic Wigner distribution is given by
\begin{equation}\label{SFFbar}
    \bar{\mathcal{P}}(t, \theta) = \sin^4 \theta + \cos^4 \theta 
    + 2 \sin^2 \theta \cos^2 \theta \, {}_1F_1\left(\frac{\alpha + 1}{2}; \frac{1}{2}; -\frac{t^2}{4A(\alpha)}\right)\,,
\end{equation}
where $A(\alpha)$ is defined in \eqref{AC}. The plateau value for the average return probability thus reads
\begin{equation}
    \bar{\mathcal{P}}(t \rightarrow \infty, \theta)
    = \frac{1}{4} (3 + \cos 4\theta) \overset{\theta = \frac{\pi}{4}}{\scalebox{2}[1]{=}} \frac{1}{2}\,.
\end{equation}
In figure \eqref{fig7}, we plot the average return probability for $\beta = 0$ ($\theta = \pi/4$) across different values of $\alpha$, and figure~\eqref{fig8} shows the return probability for varying $\theta$ in different Wigner distributions.

By comparing equations \eqref{Ckbar} and \eqref{SFFbar}, it is evident that the dips and peaks in the average spread complexity and return probability occur simultaneously. Specifically, the peak in average spread complexity corresponds to the dip in return probability, while the dip in average spread complexity aligns with the peak in average return probability, or SFF.

However, this relationship does not generally hold in a higher-dimensional RMT. As observed in several studies \cite{balasubramanian2022quantum, erdmenger2023universal, camargo2024spread}, the peak in spread complexity typically coincides with the point at which the system reaches a plateau, rather than a dip.

In RMT, particularly for the GSE, a modification at the end of the ramp is noticeable, as shown in figure \eqref{fig7}. This results in a small dip before the spread complexity saturates. For higher values of $\alpha$, the standard linear ramp is replaced by a series of dips and peaks, leading to an increased number of oscillations in spread complexity.

Let us now revisit the question raised at the beginning of this section. While the Wigner surmise is applicable to chaotic systems up to $\alpha = 4$, it is worth considering why we should not investigate larger values of $\alpha$. In the next chapter, we will show that higher values of $\alpha$ are indeed physically relevant, not only in chaotic systems but also, surprisingly, in integrable ones.
\section{Higher-Order Level Spacing in Chaotic and Integrable Systems}
\begin{figure}[h!]
 \centering
        \includegraphics[width=.6\linewidth]{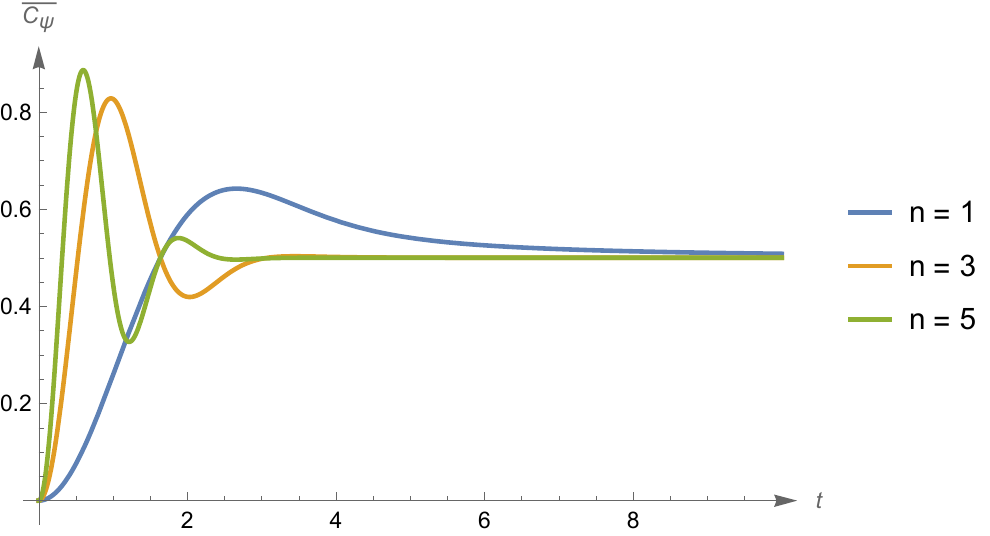}
    \caption{Average spread complexity of a superposition state in a GOE  chaotic system calculated using \eqref{average-non}. }
    \label{fig9}
\end{figure}

In many physical systems, understanding the dynamics over large spectral intervals is necessary. While the nearest neighbor level spacing reflects spectral fluctuations at the scale of the mean level spacing $\la s\ra$, higher-order level spacing can probe fluctuations over intervals as large as $n$ times the mean spacing.
Let us consider a subsystem with two levels and define the $n$th order level spacing as follows
\be
s_i^{(n)} = E_{i+n} - E_i\,,
\ee
which can be expressed as
\be
s_i^{(n)} = s_{i+n-1} + s_{i+n-2} + \ldots + s_i = \sum_{k=0}^{n-1} s_{i+k}\,.
\ee
Here, $s_i$ refers to the nearest neighbor level spacing as before. Consequently, the average spread complexity in a chaotic system characterized by the Wigner-Dyson distribution is given by
\be\label{average-non}
\bar{C}_{\psi}(t, \theta) = \int P_\alpha(s_i) P_\alpha(s_{i+1}) \ldots P_\alpha(s_{i+n-1}) \sin^2 \left(\frac{t}{2} \sum_{k=0}^{n-1} s_{i+k}\right) \sin^2(2\theta) \,.
\ee
In figure \eqref{fig9}, we observe the average spread complexity of a maximally mixed state for various values of n. It can be seen that the first peak becomes sharper, and new peaks and dips appear, exhibiting behavior similar to that of a generic Wigner distribution at higher values of $\alpha$.

Therefore, it is natural to explore the potential relationship between the level spacing distribution of energy levels $E_{i}$ and $E_{i+n}$ and the Wigner surmise with a higher exponent $\alpha$.

The quantity $P_n(s)$, where $s$ is now defined as $E_{i+n} - E_i$, is known as the \emph{higher-order level spacing distribution}. This concept has been studied in cases where the nearest neighbor level spacing follows the Wigner distribution, as discussed in \cite{rao2020higher}.

Assuming  
\begin{equation}  
P_1(s) = C(\alpha) s^\alpha e^{-A(\alpha) s^2},  
\end{equation}  
a detailed statistical analysis \cite{rao2020higher} reveals that
\begin{equation}\label{Pn}  
P_n(s) = C(\nu) s^\nu e^{-A(\nu) s^2},  
\end{equation} 
where
\begin{equation}\label{nu}
    \nu = \frac{n(n+1)}{2} \alpha + n -1\, ,
\end{equation}
with $C(\nu)$ and $A(\nu)$ as specified in \eqref{AC}.  

\begin{figure}[h!]
    \centering
    % First subfigure
    \begin{subfigure}[b]{0.5\textwidth}
        \centering
        \includegraphics[width=\linewidth]{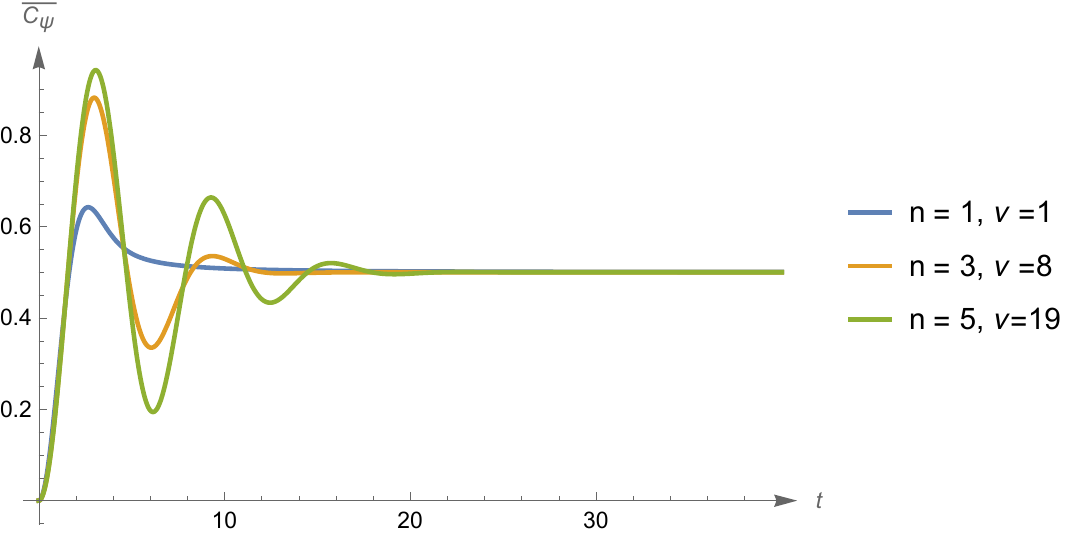}
        \caption{GOE}
        \label{fig:subfig1}
    \end{subfigure}
    \hfill
    % Second subfigure
    \begin{subfigure}[b]{0.5\textwidth}
        \centering
        \includegraphics[width=\linewidth]{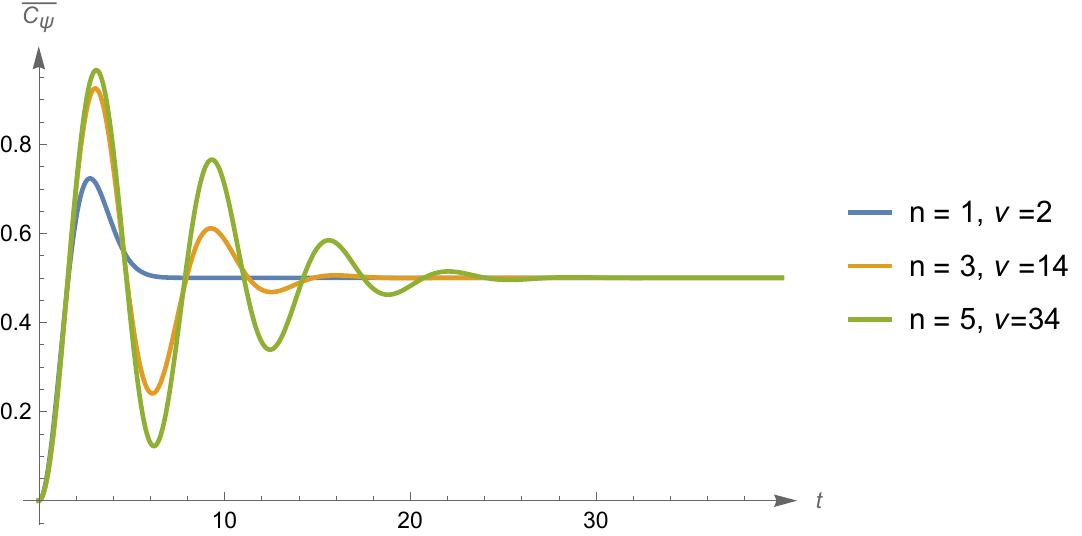}
        \caption{GUE}
        \label{fig:subfig2}
    \end{subfigure}
     % third subfigure
    \begin{subfigure}[b]{0.5\textwidth}
        \centering
        \includegraphics[width=\linewidth]{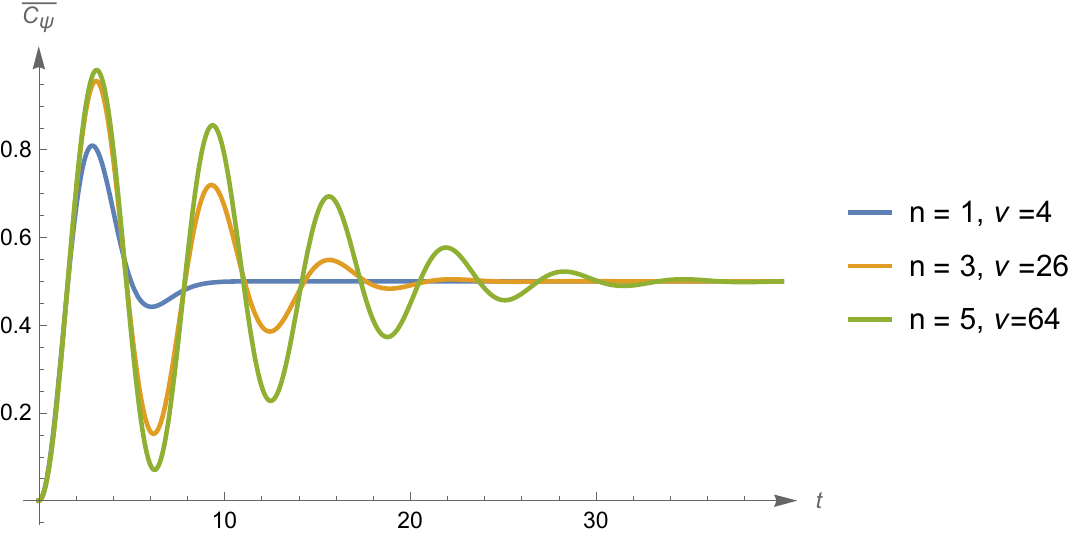}
        \caption{GSE}
        \label{fig:subfig2}
    \end{subfigure}
    \caption{Average spread complexity of the maximally mixed state for the higher level spacing distribution }
    \label{fig10}
\end{figure}

In figure~\eqref{fig10}, we plot the average spread complexity of the maximally mixed state for the level spacing distribution given in \eqref{Pn}, considering different values of $n$ in GOE, GUE, and GSE cases. As $n$ increases, the number of peaks and dips in the complexity also increases.  

It is important to note that the timing of these peaks, dips, and saturation points differs from that in the average spread complexity described by $\eqref{average-non}$. This discrepancy occurs because using $\eqref{average-non}$ is equivalent to convolving the distributions $P_{\alpha_1}(s_1)$ and $P_{\alpha_2}(s_2)$ to obtain the distribution of $P(s_1 + s_2)$. In this context, the result is not normalized as it is in $\eqref{normal}$.

Now an interesting and crucial question arises: to what extent can the characteristics of chaos remain undetected when examining higher-level spacing? To address this inquiry, it is essential to explore integrable models while considering the implications of higher-level spacing.

In an integrable system, the nearest neighbor level spacing follows a Poisson distribution, expressed as $P(s) =e^{-s}$. We can also determine the distribution of higher-order level spacings in the case of uncorrelated energy levels within the Poisson class.

Let us first consider the case where $n = 2$. We define $s = s_i + s_{i+1} = E_{i+2} - E_i$. The distribution for $s$ can be given as

\be
P(s) \propto \int_0^s ds_1\, P_1(s - s_1) P_1(s_1)= s e^{-s} \, .
\ee
Demanding the normalization condition, we find
\be
P_2(s) = 4s e^{-2s}\, .
\ee
By applying the same procedure iteratively, we can derive the distribution for higher levels of spacing within the Poisson ensemble. This leads us to
\be
P_n(s) = \frac{n^n}{(n-1)!} s^{n-1} e^{-n s}\, ,
\ee
which describes the generalized semi-Poisson distribution with index $n$.

\begin{figure}[h!]
    \centering
    \includegraphics[width=.6\linewidth]{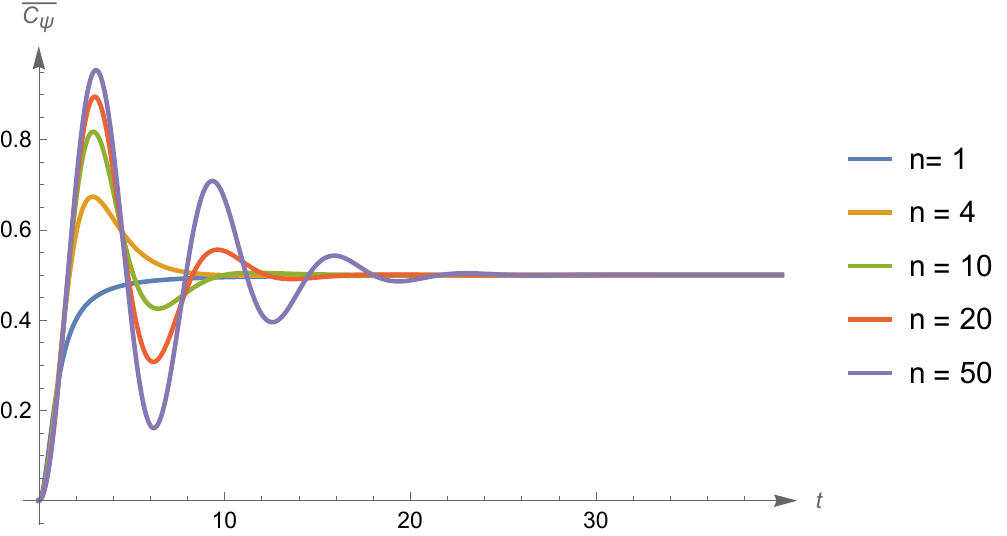}
    \caption{Average spread complexity of a two-level subsystem in an integrable system.}
    \label{fig11}
\end{figure}

\begin{figure}[h!]
    \centering
    \includegraphics[width=.6\linewidth]{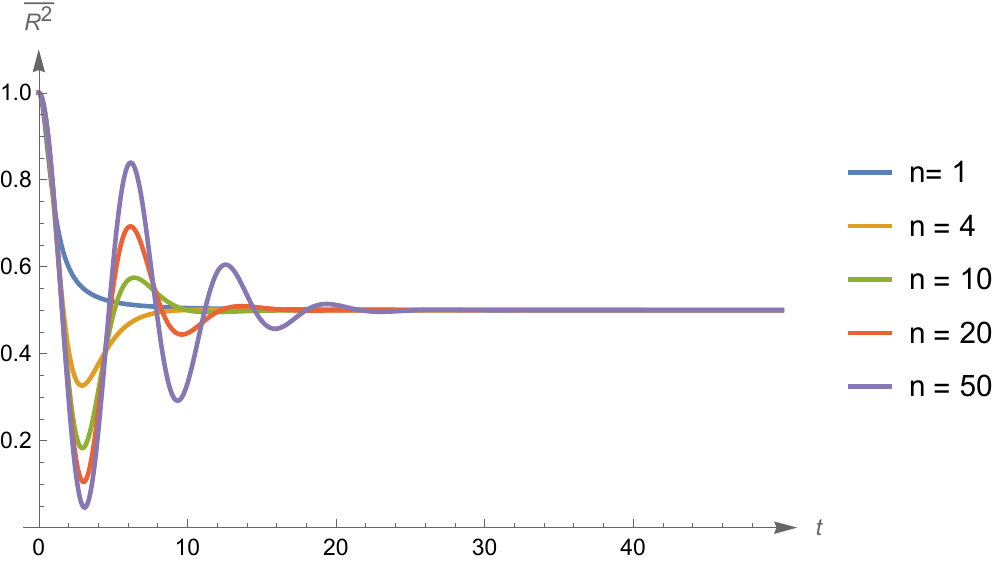}
    \caption{Average survival amplitude for \( \theta = \frac{\pi}{4} \).}
    \label{fig12}
\end{figure}

The average spread complexity for a higher level spacing reads then
\be
    \bar{C}_{\psi}(t, \theta) = & \int ds ~ P_n(s) C_\psi(t,\theta) = \frac{n^n}{4(n-1)!} \left[2 n^{-n} - (n - i t)^{-n} - (n + i t)^{-n}\right] \Gamma(n) \sin^2(2\theta)\, .
\ee
Figure \ref{fig11} shows the average spread complexity of the maximally mixed state plotted for various values of $n$.
\\
The average return probability for different semi-Poisson distributions is given by
\be
\bar{\mathcal{P}}(t, \theta)=\sin^4 \theta  + \cos^4 \theta + 2 \sin^2 \theta \cos^2 \theta \left(1 + \frac{t^2}{n^2}\right)^{-n/2} \cos\left[n \arctan\left(\frac{t}{n}\right)\right]\, .
\ee
In figure \eqref{fig12}, the results for some values of $n$ are presented. These plots, along with their qualitative agreement with observations in quantum chaotic systems, strongly support our claim. In particular, the claim is that the effect of higher order level spacing can introduce complexity that resembles quantum chaos. This observation is summarized in the conclusion section that follows.
\section{Conclusion}
In this work, we investigate the behavior of spread complexity for arbitrary quantum states, highlighting its intrinsic dependence on the projected subspace of the Hilbert space and the associated level spacing. The motivation for this approach is clear, for any initial state $\ket{\psi}$, the time evolution is confined to the subspace $\mathcal{H}_\psi$ spanned by energy eigenstates with non-zero overlap with $\ket{\psi}$. The dynamics within this subspace are governed by the projected Hamiltonian $H_\psi$, whose spectrum is a subset of the original Hamiltonian's spectrum. As a result, the spread complexity of $\ket{\psi(t)}$ is entirely determined by the level spacing of $H_\psi$. This connection establishes a powerful framework for analyzing the behavior of arbitrary quantum states in both chaotic and integrable systems. It is worth noting that the result does not depend on the explicit form of the Hamiltonian.

To proceed concretely, we analyze a two-level subsystem as a minimal model for studying spread complexity in a projected subspace. For such a state, the spread complexity depends only on the energy difference $s_{ij} = E_i - E_j$ between the two levels and exhibits simple oscillatory behavior. However, when averaged over the level spacing distribution, the spread complexity reveals richer structure. Specifically, we observe that as the energy gap widens (i.e., for higher-order level spacings), the spread complexity develops sharper peaks and additional features (more peaks and dips) before saturating. This behavior aligns with the known property that higher-order level spacings in chaotic systems display stronger correlations, which in turn induce more complex dynamical signatures in the spread complexity.

In chaotic systems characterized by Wigner-Dyson statistics, the spread complexity of a superposition state develops a characteristic pre-saturation peak that reflects the system's spectral rigidity. This peak sharpens systematically with increasing energy difference between levels, while additional oscillatory features (peaks and dips) emerge at larger energy gaps. These features originate from the progressively more structured correlations in higher-order level spacing distributions.

Notably, even integrable systems with Poissonian statistics can show similar complexity patterns when higher-order spacings are considered; however, this occurs at a slower rate than in systems that are entirely chaotic. This finding reveals that subsectors of integrable systems may display emergent chaotic-like behavior, blurring the line between integrability and chaos. This offers a fresh perspective on quantum chaos, integrability, and complexity, emphasizing the role of projected subspaces and higher-order level spacings in shaping the dynamics of quantum states.

A suggestion would be to explore an increased value of $d_\psi$ to see if it reveals a structure similar to that of $d_\psi = 2$. In this context, the projected Hamiltonian typically exhibits a spectrum that is not unique in its level spacing; rather, it reflects a variety of distributions. When the entire system is chaotic, the projected Hamiltonian tends to have an ordered spectrum as $\tilde{E}_1 < \tilde{E}_2 < \ldots < \tilde{E}_{d_\psi}$. Each level spacing $\tilde{s}_i$ conforms to a Wigner surmise distribution, characterized by its own parameter $\nu_i$, which may differ across levels. A similar behavior can be expected for integrable systems, where the distribution may follow a semi-Poisson model instead of a Wigner surmise.
To simplify this study, one might assume that all $\nu_i$ are equal and investigate whether extra peaks and dips emerge. Additionally, it would be worthwhile to determine if the number of these peaks and dips is universal for a fixed value of $\nu$, as well as the time scale required to reach these points. We will address these questions in more depth in future studies.

\section*{Acknowledgments}
We thank M. Alishahiha for carefully reading our draft and providing very useful comments.
\bibliographystyle{ieeetr}
\bibliography{refs}
\end{document}